\documentclass[twocolumn,
preprintnumbers,
tightenlines,
superscriptaddress]{revtex4-2}
\usepackage{aas_macros}
\usepackage{style}
\definecolor{darkgreen}{rgb}{0.0, 0.5, 0.0}

\begin{document}
\title{Seasons of Dark Matter Freeze-In Shaped by the Weather of the Early Universe}

\author{Francesco D'Eramo\,\orcidlink{0000-0001-8499-7685}}
\affiliation{Dipartimento di Fisica e Astronomia, Università degli Studi di Padova Via Marzolo 8, 35131 Padova, Italy}
\affiliation{Istituto Nazionale di Fisica Nucleare (INFN), Sezione di Padova, Via Marzolo 8, 35131 Padova, Italy}
	
\author{Alessandro Lenoci\,\orcidlink{0000-0002-2209-9262}}
\affiliation{Racah Institute of Physics, Hebrew University of Jerusalem, 91904 Jerusalem, Israel}
\author{Tommaso Sassi\,\orcidlink{0009-0008-8784-3655}}
\affiliation{Dipartimento di Fisica e Astronomia, Università degli Studi di Padova Via Marzolo 8, 35131 Padova, Italy}
\affiliation{Istituto Nazionale di Fisica Nucleare (INFN), Sezione di Padova, Via Marzolo 8, 35131 Padova, Italy}
	
\begin{abstract}
	Quantifying the imprints of freeze-in dark matter (DM) on cosmological structures requires knowledge of its phase-space distribution. We investigate how variations in the cosmological history before nucleosynthesis, the “weather” of that epoch, give rise to distinct “seasons” in the DM momentum distribution that governs its warmness. Studying decay-driven production across diverse cosmological histories, we map how these conditions shape DM phase-space properties. Our study quantifies how the early universe composition plays a key role in determining the mass bound on freeze-in~DM.
\end{abstract}
	
\maketitle
	
\newpage
    
\section{Introduction}
The overwhelming evidence for dark matter (DM)~\cite{Jungman:1995df,Bertone:2004pz,Feng:2010gw,Arbey:2021gdg,Cirelli:2024ssz,Bozorgnia:2024pwk} makes uncovering its particle nature a central challenge in fundamental physics. Any viable framework must reproduce the observed abundance $\Omega_{\rm DM} h^2 = 0.120 \pm 0.001$~\cite{Planck:2018vyg}, which depends on the DM microphysics and the thermal evolution of the early universe. In this context, the {\it freeze-out paradigm} remains compelling: the relic abundance naturally emerges from Hubble expansion after thermal equilibrium erases any dependence on initial conditions~\cite{Lee:1977ua,Goldberg:1983nd,Scherrer:1985zt,Srednicki:1988ce,Gondolo:1990dk}. However, the absence of signals in experimental searches~\cite{Arcadi:2017kky,Roszkowski:2017nbc,Arcadi:2024ukq} motivates the exploration of alternative scenarios.
	
The {\it freeze-in paradigm}, in which feebly coupled DM candidates never reach thermal equilibrium, is a compelling alternative. Tiny interactions can still produce DM through decays and scatterings of bath degrees of freedom. Once generated, DM follows geodesics in the expanding Friedmann–Robertson–Walker–Lemaître (FRWL) spacetime. Production via decays is dominated by temperatures near the parent mass and is insensitive to earlier history (IR domination)~\cite{Hall:2009bx}. For scatterings, IR domination occurs only for renormalizable interactions~\cite{Hall:2009bx,McDonald:2001vt,Kusenko:2006rh,Ibarra:2008kn}, while higher-dimensional operators yield UV-dominated production~\cite{Elahi:2014fsa,Chen:2017kvz,Bernal:2019mhf}. The required feeble couplings make detection extremely challenging, though several avenues have been proposed, including displaced vertices at colliders~\cite{Co:2015pka,Evans:2016zau,DEramo:2017ecx,Calibbi:2018fqf,Curtin:2018mvb,Belanger:2018sti,Junius:2019dci,No:2019gvl,Bae:2020dwf,Calibbi:2021fld,Arias:2025tvd} and direct searches~\cite{Chu:2011be,Essig:2015cda,Dvorkin:2019zdi,Boddy:2024vgt,Bernal:2024ndy}.
	
This work focuses on the cosmological imprints of freeze-in production. When DM particles are produced with substantial kinetic energy, as in decays of much heavier mother particles, they can free-stream over large distances and erase small-scale density perturbations. These effects can be probed through observables sensitive to the small-scale DM distribution, including the Lyman-$\alpha$ forest, Milky Way satellites, and strong gravitational lensing\footnote{These probes are complementary, each affected by different systematics and relying on distinct techniques.}. Quantifying the resulting power suppression requires knowledge of the DM phase-space distribution (PSD), an approach widely used to study freeze-in via decays and scatterings~\cite{Kamada:2013sh,McDonald:2015ljz,Roland:2016gli,Heeck:2017xbu,Bae:2017dpt,Boulebnane:2017fxw,Kamada:2019kpe,Dvorkin:2020xga,Ballesteros:2020adh,DEramo:2020gpr,Baumholzer:2020hvx,Egana-Ugrinovic:2021gnu,Du:2021jcj,Decant:2021mhj,Dienes:2021cxp,Xu:2024uas,Becker:2025yvb}.
	
All these studies in the literature share a {\it major caveat}: they assume that the early universe was radiation dominated. This assumption yields calculable PSDs for freeze-in via decays or scatterings through renormalizable interactions, owing to the IR domination discussed above. For scatterings mediated by higher-dimensional operators, the treatment embeds DM production within conventional reheating. In fact, this picture lacks direct observational support. The earliest signals from the primordial universe arise at the formation of the Cosmic Microwave Background (CMB), when the bath temperature was $T_{\rm CMB} \simeq 0.3\,{\rm eV}$. Earlier epochs are probed solely through Big Bang Nucleosynthesis (BBN), which constrains the thermal history back to $T_{\rm BBN} \simeq {\rm MeV}$, well below\footnote{There could be exceptions; see, e.g., Ref.~\cite{Berlin:2017ftj}.} typical freeze-in production temperatures.
	
Several studies have examined how the predicted relic density is modified when freeze-in production occurs during a non-standard cosmological era. These works analyze the evolution of the zeroth moment of the PSD (i.e., the DM number density)~\cite{Co:2015pka,Evans:2016zau,DEramo:2017ecx,Calibbi:2021fld,Bhattiprolu:2022sdd,Cosme:2023xpa,Becker:2023tvd,Silva-Malpartida:2023yks,Bernal:2024ndy,Koivunen:2024vhr,Arcadi:2024wwg,Lebedev:2024mbj,Barman:2024nhr,Silva-Malpartida:2024emu,Barman:2024tjt,Arias:2025tvd,Borah:2025ema} or the first moment~\cite{Arias:2020qty}. These calculations demonstrate that predictions vary significantly once one departs from radiation domination. This highlights how relic density calculations rest on an {\it untested assumption} about the early universe, and relaxing it alters the DM abundance.

    \begin{figure*}[!t]
	\includegraphics[width=0.98\textwidth]{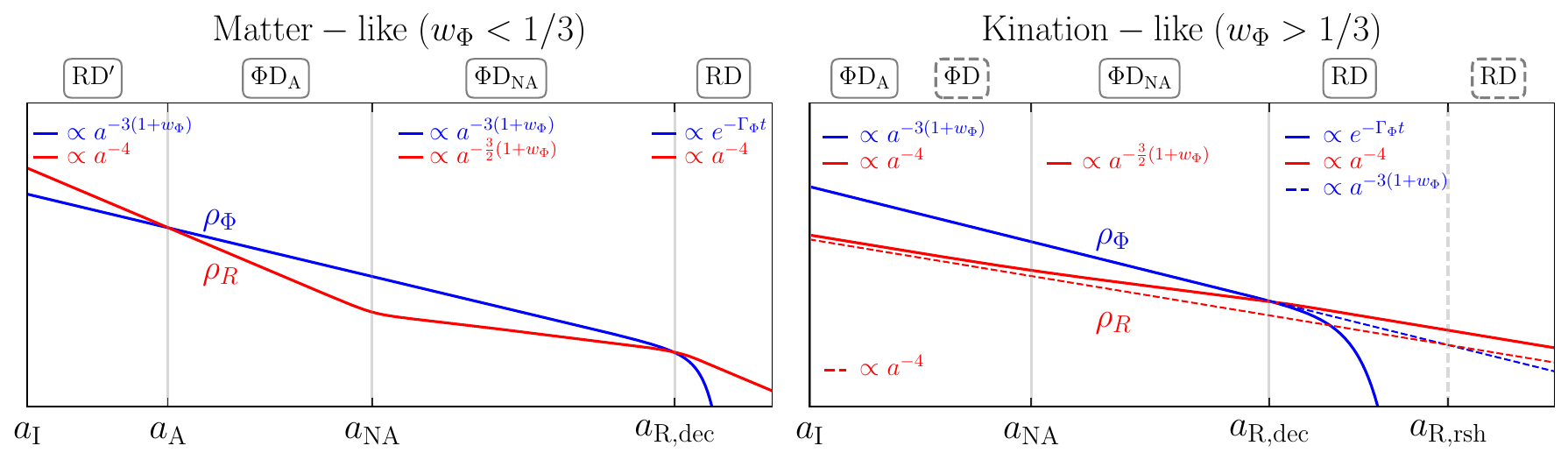}
	\caption{Cosmological histories considered in this work with a new species $\Phi$ redshifting slower (left panel) or faster (right panel) than radiation. The blue and red lines denote the energy density of $\Phi$ and radiation, respectively, both as a function of the scale factor $a$. Semi-analytical solutions for the different phases are provided in the legend. See text for further description.}
	\label{fig:histories}
    \end{figure*}

In this paper, we take this effort a step further by analyzing how non-standard cosmological histories reshape the PSD. This requires discretizing the PSD in momentum space and solving the Boltzmann equation for each bin under different cosmological backgrounds, making the approach computationally demanding. We consider non-standard histories that modify the ``{\it weather}'' of the early universe, where the primordial SM bath coexists with an additional species, and focus on production via two-body decays whose monochromatic final states enable a general treatment without specifying the underlying Lagrangian. As explicit examples show, the PSD is highly sensitive to the thermal history, producing scenarios either colder or hotter than standard freeze-in during radiation domination. These different effective temperatures lead to distinct ``{\it seasons}'' of DM freeze-in.
	
Our analysis yields concrete, testable results, providing novel lower mass bounds on DM with direct phenomenological implications. Fixing just the mother particle mass, we scan over cosmological histories to quantify their impact. The resulting bounds show that early-time cooling mechanisms can significantly relax the limits, emphasizing the strong sensitivity of DM constraints to the assumed cosmological history.

The main body of this paper investigates the physical implications of modified cosmological backgrounds for freeze-in DM and the resulting structure-formation constraints. For completeness, two appendices present semi-analytical treatments of the cosmological backgrounds considered in this study, together with solutions for the PSD. Although all results shown in the main text are obtained through numerical computations, the analytical developments in the appendices clarify the underlying dynamics and provide complementary insight that sharpens the interpretation of the numerical findings.
	
	
\section{Early Universe Weather}
We consider deviations from standard cosmology in which the early universe contains an additional species $\Phi$ with equation of state $p_\Phi = w_\Phi \rho_\Phi$. If $\Phi$ is unstable, it decays with width $\Gamma_\Phi$, and we assume its decay products thermalize rapidly with the primordial plasma. The evolution of the energy densities of $\Phi$ and of the radiation bath is governed by
	\begin{subequations}\label{eq:bkg}
		\begin{align}
			& \frac{d\rho_\Phi}{dt} + 3(1 + w_\Phi) H \rho_\Phi = -\Gamma_\Phi \rho_\Phi \ , \\ 
			& \frac{d\rho_R}{dt} + 3(1 + w_R) H \rho_R = \Gamma_\Phi \rho_\Phi \ .
		\end{align}
	\end{subequations}
	The Hubble parameter, defined as $H \equiv \dot{a}/a$ with $a(t)$ the FRWL scale factor, obeys the Friedmann equation $3H^2 M_{\rm Pl}^2 = \rho_\Phi + \rho_R$~\footnote{We employ the reduced Planck mass defined as $M_{\rm Pl} \equiv (8\pi G)^{-1/2} \simeq 2.4 \times 10^{18}\,{\rm GeV}$.}. First, we neglect the subdominant DM energy density and determine the cosmological background, and then study DM production within it.
	
	We use the scale factor $a$ as the evolution variable. Initial conditions for $\rho_\Phi$ and $\rho_R$ are set at an arbitrary initial scale factor value $a_{\rm I}$, with the corresponding bath temperature $T_{\rm I}$ determined from $\rho_R(T_{\rm I}) = (\pi^2/30)\, g_\star(T_{\rm I})\, T_{\rm I}^4$. For the effective degrees of freedom $g_\star(T)$ and for $w_R(T)$, we use the results of Ref.~\cite{Laine:2015kra}.
	
	The plots in Fig.~\ref{fig:histories} illustrate how the energy densities evolve. In the left panel, we show the {\it matter-like}~\footnote{Non-relativistic matter, $w_\Phi = 0$, is a case of this kind.} case satisfying $w_\Phi < 1/3$. Initially, the energy density of $\Phi$ can be larger or smaller than that of radiation, but its different redshift behavior eventually makes radiation subdominant, requiring $\Phi$ to be unstable. The plot shows the case where $\Phi$ is initially subdominant, so the universe first undergoes a radiation-dominated phase (RD') for scale factor values in the range $(a_{\rm I}, a_{\rm A})$. During the first adiabatic stage of $\Phi$ domination, $(a_{\rm A}, a_{\rm NA})$, denoted $\Phi\text{D}_{\rm A}$, the comoving entropy of the radiation bath is conserved. As $\Phi$ decays, the radiation bath becomes dominated by its decay products, leading to a non-adiabatic phase, $(a_{\rm NA}, a_{\rm R,dec})$, denoted $\Phi\text{D}_{\rm NA}$, during which entropy is not conserved. This phase continues until $\Phi$ fully decays at $a_{\rm R,dec}$, after which the standard radiation-dominated universe (RD) is recovered. The right panel of Fig.~\ref{fig:histories} illustrates the {\it kination-like}~\footnote{Kination, $w_\Phi = 1$, is a case of this kind.} scenario satisfying $w_\Phi > 1/3$. We consider unstable $\Phi$ (solid lines) and stable $\Phi$ (dashed lines). To have any impact, the energy density of $\Phi$ must dominate at the initial time $a_{\rm I}$, and the first phase $(a_{\rm I}, a_{\rm NA})$ is adiabatic. Eventually, the decay products become sufficient to trigger a non-adiabatic phase. The radiation-dominated universe is recovered either when $\Phi$ decays at $a_{\rm R,dec}$ or redshifts away at $a_{\rm R,rsh}$.
	
	\vspace{0.1cm}
	
	\section{Freeze-in Seasons}
    We consider DM particles $\chi$ produced via two-body decays $\mathcal{B}_1 \rightarrow \mathcal{B}_2 \chi$ of a bath particle $\mathcal{B}_1$ with mass $M$. For decays yielding two DM particles, the conclusions are unchanged apart from a factor of two in the lifetime needed to reproduce the relic density. We take the bath particle $\mathcal{B}_2$ to be massless, as a finite mass introduces a known phase-space suppression~\cite{DEramo:2020gpr}. We also assume $M \gg m_\chi$, justified since the mass bounds derived here lie in the tenths-of-keV range.
	
	Homogeneity and isotropy restrict the PSD to depend only on FRWL time $t$ and the physical-momentum magnitude $p(t)$. The PSD obeys the Boltzmann equation
	\begin{equation}\label{eq:Boltzmann}
		\frac{d f_\chi(t,p(t))}{dt} = {\cal C}_{\mathcal{B}_1 \rightarrow \mathcal{B}_2 \chi} (T(t), p(t))\ .
	\end{equation}
	We adopt Maxwell–Boltzmann (MB) statistics for all particles, since quantum degeneracy has a negligible impact on the DM PSD~\cite{DEramo:2020gpr,DEramo:2025jsb}. Under these assumptions, the collision term is known analytically~\cite{DEramo:2020gpr}
	\begin{align}
		{\cal C}_{\mathcal{B}_1 \rightarrow \mathcal{B}_2 \chi}(T,p) = g_{{\cal B}_{1}}\Gamma_{{\cal B}_{1}} \frac{MT}{p^2}\exp\left(-\frac{p}{T}-\frac{M^2}{4p T}\right)\ ,
		\label{eq:CollisionTerm2Bdec}
	\end{align}
	where $g_{{\cal B}_{1}}$ and $\Gamma_{{\cal B}_{1}}$ denote the internal degrees of freedom and decay width of the mother particle, respectively. Formally, the Boltzmann equation has the following solution
	\begin{align}\label{eq:f_sol_general}
		f_\chi(a,p(a)) = \int_{a_{\rm I}}^a \frac{da'}{a'}\   \frac{{{\cal C}_{\mathcal{B}_1 \rightarrow \mathcal{B}_2 \chi}(T(a'),p(a'))}}{H(a')} \ .
	\end{align}
	For a given cosmological background, one can evaluate $H(a')$ and $T(a')$ and compute the integral in Eq.~\eqref{eq:f_sol_general} numerically to obtain the PSD at any time. As a self-consistency check of the freeze-in assumption, it is crucial to verify that the DM number density always remains well below equilibrium, $n_\chi \ll n_\chi^{\rm eq} \sim T^3$. We require $n_\chi$ to be always one order of magnitude below $n_\chi^{\rm eq}$. Evaluating the PSD at $a = a_0$, where the subscript $0$ denotes quantities today, yields the present value $f_\chi(a_0,p_0)$. This allows one to determine the value of $g_{{\cal B}_1}\Gamma_{{\cal B}_1}/M$ required to reproduce the observed DM relic density.
	
	A key quantity is the second moment of the PSD, which sets the root-mean-square (r.m.s.) DM velocity, $W_\chi \equiv \sqrt{\langle p_0^2 \rangle}/m_\chi$. Its pivotal role in cosmic structure formation was highlighted by Refs.~\cite{Kamada:2013sh,McDonald:2015ljz,Roland:2016gli,Heeck:2017xbu,Kamada:2019kpe} and confirmed quantitatively in Ref.~\cite{DEramo:2020gpr}.
	
	We solve the Boltzmann equation in terms of the comoving momentum $q \equiv p\,a/(M a_M)$, which factors out the expansion, with $a_M$ defined by $T(a_M)=M$. While convenient for computing the PSD, caution is required when connecting $q$ to structure formation. During radiation domination, the second moment $\sigma_q \equiv \sqrt{\langle q^2 \rangle}$ remains constant once production ends, but in modified cosmologies entropy release during or after DM production spoils this relation. 
    
    A more robust diagnostic quantity is
	\begin{equation}
		\Sigma \equiv \frac{\sqrt{\langle p_0^2\rangle}}{T_\chi(t_0)} \ ,
		\label{eq:Ttilde}
	\end{equation}
	where $T_\chi(t_0) = T_0 [g_{\star s}(T_0)/g_{\star s}(M)]^{1/3}$ defines the DM effective temperature today. It is related to the present r.m.s. DM velocity through $\Sigma = m_\chi W_\chi(t_0)/T_\chi(t_0)$, and we will use it to set bounds on the DM mass. To relate the comoving quantity $\sigma_q$ to its physical counterpart $\Sigma$, we introduce the ratio between the entropy in a comoving volume at the end of the non-adiabatic phase and at the scale factor value $a_M$, $D(M) = S(a_{\rm R,dec}) / S(a_M)$. This quantity is the dilution factor generated by the entropy injected in the plasma. The relation between $\Sigma$ and $\sigma_q$ is $\Sigma = D(M)^{-1/3} \sigma_q$. As a check, for $M < T_{\rm R}\equiv T(a_{\rm R,dec})$ or in the absence of a non-adiabatic phase, $D(M) = 1$, and $p_0 / T_\chi(t_0) = q\, D(M)^{-1/3}$ correctly gives $\Sigma = \sigma_q$.
	
	\begin{figure}
		\centering
		\includegraphics[width=\linewidth]{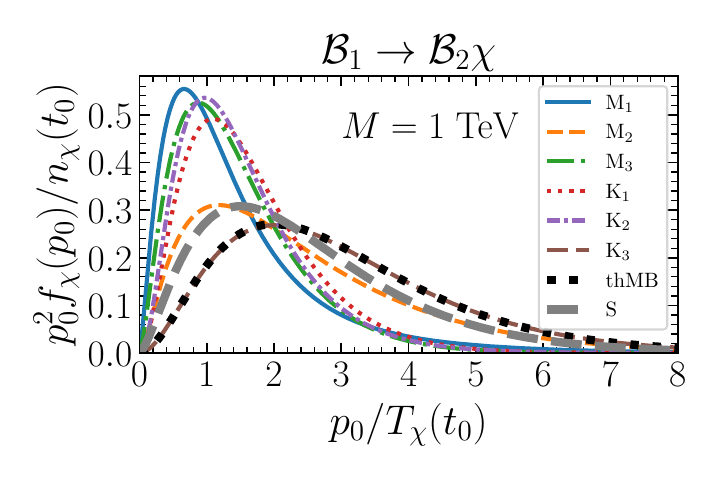}
		\caption{Numerical solutions for $(M, m_\chi) = (1\,\mathrm{TeV}, 30\,\mathrm{keV})$ with $g_{{\cal B}_1}\Gamma_{{\cal B}_1}$ fixed to reproduce the observed DM relic density. PSDs, normalized to the present relic density, are shown as functions of the normalized DM momentum $p_0 / T_\chi(t_0)$. Colored lines denote the benchmark scenarios in Tab.~\ref{tab:solutions}, and the black dotted line shows a MB thermal PSD.}
		\label{fig:PSDs}
	\end{figure}
	
	Fig.~\ref{fig:PSDs} shows numerical solutions of the Boltzmann equation for $(M, m_\chi) = (1\,\mathrm{TeV}, 30\,\mathrm{keV})$, with the mother decay width fixed by the relic density requirement. Each color corresponds to a different cosmological history for freeze-in production, while the thick dotted black line shows a thermal MB distribution for reference. The PSDs are plotted as functions of the present-day normalized physical momentum $p_0/T_\chi(t_0)$, after any entropy release in the early universe, and normalized to the relic number density. The figure illustrates that, for fixed particle physics, the cosmological history markedly reshapes the distributions, yielding \textit{hotter} or \textit{colder} DM temperatures.

    \begin{table*}[ht]
		\begin{center}
			{\def\arraystretch{1.3}
				\begin{tabular}{c||ccc|c|c|c|c|c|c|c}
					& $\rho_\Phi(a_{\rm I})[{\rm GeV}^4]$ & $\rho_R(a_{\rm I})[{\rm GeV}^4]$ & $\Gamma_\Phi[{\rm GeV}]$ & phase & $(g_{{\cal B}_1}\Gamma_{{\cal B}_1}/M)_{\rm relic}$ & $\sigma_q$ & $D(M)$ & $\Sigma$ & $\Sigma_{\rm th}$ & $(\alpha,\,\beta,\,\gamma)$ \\
					\hline\hline
					S & $0$ & $3.5\times10^{49}$ & $0$ & RD &$2.0\times10^{-16}$ & 3.0 & 1 & $3.0$ & $3.0$ & $(-0.5,1.0,1.0)$ \\
					\hline
					\textcolor[HTML]{1f77b4}{ $\rm M_1$} & $3.2\times10^{32}$ & $5.5\times10^{38}$ & $2.0\times10^{-14}$ & $\Phi {\rm D}_{\textup{A}}/$$\Phi {\rm D}_{\textup{NA}}$ & $3.1\times10^{-15}$ & $4.3$ & $11$ & $1.9$ & $1.2$ & $(-0.4,3.1,0.7)$ \\
					\textcolor[HTML]{ff7f0f}{$\rm M_2$} & $1.0\times10^{57}$ & $3.5\times10^{61}$ & $1.2\times10^{-14}$ & $\Phi {\rm D}_{\textup{NA}}$ & $9.6\times10^{-14}$ & $300$ & $10^6$ & $3.0$ & 3.8 & $(-0.1,2.7,0.6)$ \\
					\textcolor[HTML]{2ba02b}{$\rm M_3$} & $1.0\times10^{30}$ & $1.8\times10^{38}$ & $5.5\times10^{-18}$ & $\mathrm{RD}^\prime$ & $1.0\times10^{-15}$ & $3.0$ & $5.0$ & $1.7$ & $1.6$ & $(-0.5,1.7,1.0)$ \\
					\hline
					\textcolor[HTML]{d62727}{$\rm K_1$} & $1.0\times10^{35}$ & $3.4\times10^{21}$ & $2.0\times10^{-20}$ & $\Phi{\rm{D}}_{\textup{A}}$ & $1.9\times10^{-11}$ & $3.5$ & $6.1$ & $1.9$ & $1.9$ & $(0.0,1.9,1.0)$\\
					\textcolor[HTML]{9467bd}{$\rm K_2$} & $1.0\times10^{35}$ & $3.4\times10^{21}$ & $1.0\times 10^{-16}$ & $\Phi {\rm D}_{\textup{NA}}$ & $3.2\times10^{-12}$ & $8.7$ & $110$ & $1.8$ & 1.8  & $(0.5,3.1,0.8)$ \\
					\textcolor[HTML]{8c564c}{$\rm K_3$} & $6.3\times10^{41}$ & $2.2\times10^{28}$ & $0$ & $\Phi {\rm D}$ & $6.3\times10^{-14}$ & $3.5$ & 1 & $3.5$ & 3.5 & $(0.0,1.0,1.0)$ 
				\end{tabular}
			}
		\end{center}
		\caption{Benchmark scenarios for the cosmological seasons of freeze-in production via two-body decays, corresponding to the numerical results in Fig.~\ref{fig:PSDs}. The first row refers to standard cosmology, the next three to matter domination ($w_\Phi = 0$), and the last three to kination ($w_\Phi = 1$). For each case, we provide the initial conditions $(\rho_\Phi(a_{\rm I}),\,\rho_R(a_{\rm I}))$, the decay width $\Gamma_\Phi$, the phase of freeze-in production, the $g_{{\cal B}_1}\Gamma_{{\cal B}_1}/M$ needed to reproduce the DM relic density for $m_\chi = 30\,\keV$. We also report the main PSD features, including the second moment in comoving and physical variables and the fit parameters of Eq.~\eqref{eq:PSDfit}.}
		\label{tab:solutions}
	\end{table*}
	
	Tab.~\ref{tab:solutions} summarizes the quantitative properties of the cosmological histories underlying the scenarios in Fig.~\ref{fig:PSDs}. The warmness of the PSD is quantified by fitting it to
	\begin{equation}\label{eq:PSDfit}
		f_\textup{fit}(P) \propto P^{\,\alpha} e^{-\beta P^{\,\gamma}}\,,
	\end{equation}
	where $P \equiv p_0/T_\chi(t_0)$. The last column of Tab.~\ref{tab:solutions} lists the best-fit parameters for each case. The coefficients $(\alpha,\gamma)$ are determined mainly by the cosmological phase during freeze-in production, with small corrections from residual production in nearby phases. By contrast, $\beta$ depends sensitively on the post-production history, while the combination $\beta_q \equiv \beta\, D(M)^{-\gamma/3}$ remains invariant.
	
	The behavior of the PSD can be captured analytically, and these estimates also yield analytical calculations of the second moment reported as $\Sigma_{\rm th}$ in Tab.~\ref{tab:solutions}. Assuming that most DM production occurs within a single cosmological phase, the temperature scales as $T = M\,(a/a_M)^\ell$, with $\ell=-1$ for adiabatic and $\ell=-3(1+w_\Phi)/8$ for non-adiabatic evolution. The Hubble rate follows a power law, $H \propto a^m$, with $m=-2$ for radiation domination and $m=-3(1+w_\Phi)/2$ for $\Phi$ domination. The PSD in comoving momentum can then be expressed as
	\begin{align}\label{eq:PSD_an}
		f_\chi(a,q) \propto \frac{1}{q^2}
		\int_{a_{\rm I}/a_M}^{a/a_M} \! dx\; x^{\ell-m+1}
		\exp\!\left[-q\,x^{-\ell-1}-\frac{x^{1-\ell}}{4q}\right] \ .
	\end{align}
	We omit the overall normalization since we focus on the shape. In some cases, this integral can be evaluated analytically. The PSD second moment at a given stage of the expansion identified by the scale factor $a$ results in
	\begin{align}\label{eq:sigmaq_an}
		\sigma_q\!\left(\frac{a}{a_M}\right)
		= \frac{1}{2}
		\sqrt{
			\frac{
				\displaystyle\int_{a_{\rm I}/a_M}^{a/a_M} \! dx\; x^{\ell-m+4} K_3(x^{-\ell})
			}{
				\displaystyle\int_{a_{\rm I}/a_M}^{a/a_M} \! dx\; x^{\ell-m+2} K_1(x^{-\ell})
		}}\,,
	\end{align}
	with $K_{1,3}$ modified Bessel functions. We need the present-day second moment. For adiabatic phases ($\ell = -1$), we can take the limits $a/a_M \to \infty$ and $a_{\rm I}/a_M \to 0$ in the above expression. In contrast, the $a/a_M \to \infty$ limit is not allowed for DM production during non-adiabatic phases, where production happens while the bath is being injected with entropy from $\Phi$ decays. The appropriate limit for freeze-in during non-adiabatic phases will be given approximately by the duration of the non-adiabatic phase from $a_M$ until reheating $a_{\rm R,dec}$, i.e. $a/a_M\to a_{\rm R,dec}/a_M = D(M)^{1/(3+3\ell)}$. The results $\sigma^{\rm A}_q=\sigma_q(\infty)$ and $\sigma_q^{\rm NA}=\sigma_q(a_{\rm R,dec}/a_M)$ allow us to estimate $\Sigma$ and derive analytical DM lower bounds on the DM mass.
	
	\vspace{0.1cm}
	
	\section{Mass Bounds from Structures}
    We use the PSDs obtained in modified cosmologies to derive DM mass bounds. Following Refs.~\cite{Kamada:2019kpe,DEramo:2020gpr,DEramo:2025jsb}, we compare the DM warmness $W_\chi$ with the reference warm dark matter (WDM) limit $W^{\rm max}_{\rm WDM}$, imposing $W_\chi(t_0) < W^{\rm max}_{\rm WDM}(t_0)$. This condition sets the minimum allowed DM mass
	\begin{align}
		m_\chi > \frac{\sqrt{\langle p_0^2\rangle}}{W^{\rm max}_{\rm WDM}(t_0)} 
		= \frac{\Sigma\, T_\chi(t_0)}{W^{\rm max}_{\rm WDM}(t_0)}\,.
	\end{align}
	
	Due to observational and modeling systematics in small-scale WDM probes, complementary constraints are essential. The most stringent limits are provided by Milky Way satellite counts, strong lensing, and the Lyman-$\alpha$ forest, as summarized in Fig.~2 of Ref.~\cite{DEramo:2025jsb}. Using the explicit expression for $W^{\rm max}_{\rm WDM}$\footnote{See App.~A of Ref.~\cite{DEramo:2025jsb} for a detailed derivation.}, we obtain
	\begin{align}\label{eq:m_min}
		m_\chi^{\rm min} = 19\,{\rm keV} 
		\left(\frac{m^{\rm min}_{\rm WDM}}{6\,\rm keV}\right)^{4/3}
		\left(\frac{\Sigma}{3}\right)
		\left(\frac{104.4}{g_{\star s}(M)}\right)^{1/3}\,.
	\end{align}
	Here, we adopt $m^{\rm min}_{\rm WDM} = 6\,\mathrm{keV}$ as our benchmark. This result mirrors the bounds derived in Ref.~\cite{DEramo:2025jsb}, with the crucial difference that $\sigma_q$ is replaced by $\Sigma$ to account for the impact of the cosmological background. Enforcing both the relic density constraint and freeze-in consistency, $D(M)\lesssim 10\, n_\chi(t_0)/n_\chi^{\rm eq}$, implies $m_\chi \gtrsim 2\,{\rm keV}\, D(M)$, which can compete with $m_\chi^{\rm min}$. This requirement does not apply to production during non-adiabatic phases or during transitions between eras—such as in $\mathrm{M_1}$, $\mathrm{M_2}$, and $\mathrm{K_2}$—where a full numerical treatment is necessary. Tab.~\ref{tab:solutions_2} summarizes the DM mass bounds, obtained numerically and estimated analytically, for the benchmarks in Tab.~\ref{tab:solutions} and Fig.~\ref{fig:PSDs}.
	
	\begin{table}[t]
		\def\arraystretch{1.5}
		\begin{tabular}{c||c|c|c|c|c|c|c}
			& $\rm S$ &    \textcolor[HTML]{1f77b4}{ $\rm M_1$} &   \textcolor[HTML]{ff7f0f}{$\rm M_2$} &   \textcolor[HTML]{2ba02b}{$\rm M_3$}&   \textcolor[HTML]{d62727}{$\rm K_1$}&   \textcolor[HTML]{9467bd}{$\rm K_2$}&   \textcolor[HTML]{8c564c}{$\rm K_3$}\\
			\hline \hline
			numerical  $m_\chi ^{\rm min}$ [keV]&  $19$ & $12$ & $19$ & $11$ & $12$ & $11$ & $22$\\
			\hline
			analytical $m_\chi ^{\rm min}$ [keV]&  $19$ & $7.6$ & $24$ & $10$ & $12$ & $11$ & $22$\\
		\end{tabular}
		\caption{Lower mass bounds $m_\chi^{\rm min}$ from Eq.~\eqref{eq:m_min} for the benchmark scenarios in Tab.~\ref{tab:solutions}. Both the numerical results and the bounds derived from the analytical estimate in Eq.~\eqref{eq:sigmaq_an}, combined with the appropriate dilution factor, are shown.}
		\label{tab:solutions_2}
	\end{table}
	
	\vspace{0.1cm}
	
\section{Conclusions}
The freeze-in paradigm remains a compelling scenario for DM production. Although the required feeble interactions make experimental tests challenging, recent years have brought promising strategies. Cosmological observations play a key role: they can reveal imprints in small-scale structure or, when no deviations are seen, set lower mass bounds with implications for collider and direct-detection searches.
	
In this paper, we revisited cosmological mass bounds beyond the standard assumption of radiation domination, which, though plausible, lacks direct observational support. This required a greater computational effort than conventional freeze-in analyses, as we solved the full Boltzmann equation in phase space. Our approach relies on a practical framework with targeted simplifying assumptions along two complementary directions. First, we modeled the cosmological history macroscopically, treating the exotic component $\Phi$ as a fluid without specifying its microscopic origin. Second, we focused on DM production through two-body decays, whose monochromatic final states allow for a general analysis.
	
\begin{figure}[t]
\centering
\includegraphics[width=1\linewidth]{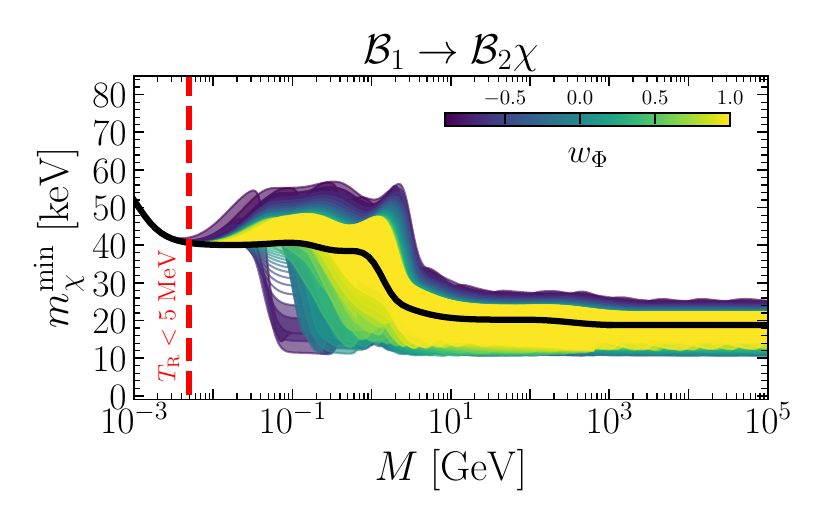}
\caption{Minimum allowed DM mass as a function of the parent mass $M$ for two-body decays.  The black curve corresponds to the standard cosmological history, while the color coding indicates different values of the equation-of-state parameter $w_\Phi \in [-0.9, 1]$.}
	\label{fig:scan}
\end{figure}

We found that the DM mass bound can vary by about a factor of two, motivating a broader exploration of the cosmological parameter space. Fig.~\ref{fig:scan} addresses this by showing quantitatively how the bound varies with different cosmological histories for $M$ in the range between 1 MeV and 100 TeV. We vary the equation-of-state parameter in the range $w_{\Phi}\in [-0.9,1]$ and explore the different parameters $(\rho_\Phi(a_{\rm I}),\rho_R(a_{\rm I}),\Gamma_\Phi)$. In our scan, we consider two distinct cosmological backgrounds that yield 
the same dilution factor as equivalent; indeed, the dilution factor is the relevant quantity determining $m_\chi^{\rm min}$. The initial radiation temperature is always chosen to satisfy $T_{\rm I} \gtrsim 10\,M$ in order to keep the mother particle in the bath. We further impose $T_{\rm R} > 5\,\mathrm{MeV}$ 
\cite{Hannestad:2004px,deSalas:2015glj,Barbieri:2025moq}.  Fig.~\ref{fig:scan} shows the resulting lower mass bound as a function of the decaying particle mass $M$, with the color coding indicating different values of $w_{\Phi}$. We observe that alternative cosmological histories can either relax or strengthen the structure-formation bound obtained in the 
standard radiation-dominated scenario~\cite{DEramo:2025jsb}, shown in black. The lower bound is dictated by the self-consistency of the freeze-in mechanism: for lighter DM masses, inverse processes become non-negligible and the system would approach thermalization, thereby invalidating the freeze-in assumption. 

We emphasize that the methodology developed in 
Refs.~\cite{Kamada:2019kpe,DEramo:2020gpr,DEramo:2025jsb} proves particularly powerful and efficient in this context. A direct scan over cosmological histories, determining the mass bound by explicitly solving the full Boltzmann system with numerical codes such as \texttt{CLASS}~\cite{Lesgourgues:2011rh} for each cosmological background, would be significantly more 
computationally demanding.
	
Beyond yielding phenomenologically meaningful results, our analysis opens several promising directions for future research. A natural extension is to move beyond decay-driven production and investigate single or double production through scatterings, where additional subtleties arise from the interplay between IR and UV domination. Moreover, DM may constitute only a subcomponent of the total abundance, in which case the corresponding mass bounds would be relaxed. Extending phase-space studies beyond DM—for instance, to dark radiation~\cite{DEramo:2023nzt,DEramo:2024jhn,Badziak:2024qjg}—within modified cosmological histories would also be of interest. In a complementary direction, implementing explicit microscopic models of reheating, kination, or other nonstandard cosmologies would provide further insight. Finally, since our analysis relies on rescaling methods akin to those used for WDM, developing an analogous framework for modified cosmologies, in the spirit of Ref.~\cite{DEramo:2020gpr}, represents a valuable next step. We leave these developments to future work.
	
    
\acknowledgements F.D. and T.S. are supported by Istituto Nazionale di Fisica Nucleare (INFN) through the Theoretical Astroparticle Physics (TAsP) project, and in part by the Italian MUR Departments of Excellence grant 2023-2027 ``Quantum Frontiers''. The work of T.S. is supported in part by the Italian Ministry of University and Research (MUR) through the PRIN 2022 project n. 20228WHTYC (CUP:I53C24002320006 and C53C24000760006). A.L is grateful to the Azrieli Foundation for the award of an Azrieli Fellowship. A.L is supported by an ERC STG grant (``Light-Dark,'' grant No. 101040019). 
This project has received funding from the European Research Council (ERC) under the European Union’s Horizon Europe research and innovation programme (grant agreement No. 101040019).  Views and opinions expressed are however those of the author(s) only and do not necessarily reflect those of the European Union. The European Union cannot be held responsible for them. This article is based upon work from COST Action COSMIC WISPers CA21106, supported by COST (European Cooperation in Science and Technology).

\appendix

	
\section{Cosmological Histories}\label{supp:cosmo}
	
In this Appendix, we momentarily set aside the specifics of DM production and concentrate on the cosmological background in which it unfolds. We consider an early universe composed of two components: the standard radiation bath with energy density $\rho_R$, and an additional exotic sector $\Phi$ with energy density $\rho_\Phi$. Their evolution follows the coupled Boltzmann equations reported in Eq.~\eqref{eq:bkg}, which also defines $w_\Phi$ and $\Gamma_\Phi$.
	
\subsection{Boltzmann system for the cosmological background in dimensionless and comoving variables}
\label{supp:cosmoBE}

We begin by recasting the system in a more convenient form by introducing dimensionless variables. In particular, we replace the cosmic time $t$ with the Friedmann–Robertson–Walker–Lemaître (FRWL) scale factor $a$. The Boltzmann evolution is assumed to start at an initial epoch characterized by $a_{\rm I}$. Since only ratios of scale factors are relevant for the subsequent dynamics, we define the reduced variable $A \equiv a/a_{\rm I}$, such that the evolution begins at $A_{\rm I}=1$. It is then natural to introduce the comoving, dimensionless energy densities
	\begin{subequations}
		\begin{align}
			\label{eq:Phi_comoving} 
			F(A) &\equiv \frac{\rho_\Phi(A)}{\overline T^4}\, A^{3(1+w_\Phi)}, \\  
			\label{eq:R_comoving} 
			R(A) &\equiv \frac{\rho_R(A)}{\overline T^4}\, A^{4}
			= \frac{\pi^2 g_\star(T{(A)})}{30}\, \frac{T(A)^4}{\overline T^4}\, A^{4}\, .
		\end{align}
	\end{subequations}
Here $\overline T$ is an arbitrary reference temperature introduced to render the above quantities dimensionless. The explicit $A$-dependence in Eqs.~\eqref{eq:Phi_comoving}–\eqref{eq:R_comoving} is chosen such that $F(A)$ is unaffected by Hubble dilution, while for $R(A)$ the impact of expansion is minimized (up to $g_\star(T)$ corrections) by matching the redshift behavior of an ultrarelativistic fluid. In the second equality of Eq.~\eqref{eq:R_comoving}, we have used the standard relation for $\rho_R(T)$ in terms of the effective number of relativistic degrees of freedom $g_\star(T)$.
	
A convenient choice is to take $\overline T$ as the temperature of the radiation bath $T_{\rm I}$ at the initial time $A_{\rm I}$. In that case, one simply has $R_{\rm I} \equiv R(A_{\rm I}) = (\pi^2/30)\, g_\star(T_{\rm I})$. This choice is, of course, not applicable if no radiation bath is present initially, as in inflationary reheating. In all scenarios considered here, however, DM production during reheating is equivalent to production in the $\rm \Phi D_{NA}$ phase shown in the left panel of Fig.~\ref{fig:histories}. Adopting $\overline T = T_{\rm I}$ therefore entails no loss of generality.
	
With these variables in hand, the Boltzmann system in Eq.~\eqref{eq:bkg} can be rewritten as
	\begin{subequations}
		\begin{align}
			\label{eq:bkg_comovingF}   
			\frac{dF(A)}{d\log A} &= -\frac{\Gamma_\Phi}{H(A)}\,F(A)\,, \\
			\label{eq:bkg_comovingR}   
			\frac{dR(A)}{d\log A} &= \bigl(1 - 3 w_R(A)\bigr)\,R(A) 
			+ \frac{\Gamma_\Phi}{H(A)} F(A)\,A^{(1 - 3 w_\Phi)} \ .
		\end{align}
		\label{eq:bkg_comoving}
	\end{subequations}
The Hubble parameter appearing in both equations follows from the Friedmann equation
	\begin{equation}
		H(A) = \frac{\sqrt{F(A)\,A^{1 - 3 w_\Phi} + R(A)}}{\sqrt{3}\,A^{2}}\,
		\frac{T_{\rm I}^2}{\Mpl}\,.
		\label{eq:H_comoving}
	\end{equation}
	
This reformulation of the Boltzmann system, which recasts the dynamics in a more convenient set of variables, is also central to our numerical analysis. Working with these variables substantially streamlines the numerical integration and forms the basis of the computational framework used to obtain all results presented in this paper. In the following, we build on this setup by presenting semi-analytical solutions and identifying the key phases of the cosmological evolution. For simplicity of exposition, we neglect the mild temperature dependence of $g_\star(T)$ and set $w_R(T)=1/3$.
	
\subsection{Semi-analytical solutions for matter-like $\Phi$}
\label{supp:semianalytic_cosmo}
	
	We investigate the cosmological history in scenarios where the primordial thermal bath is supplemented by an exotic fluid with equation-of-state parameter $w_\Phi < 1/3$. A representative solution of the Boltzmann system in this regime is shown in the left panel of Fig.~\ref{fig:histories}, where the main phases of the evolution are highlighted. Motivated by this behavior, we adopt initial conditions satisfying $F_{\rm I} \equiv F(A_{\rm I}) < R_{\rm I}$. This assumption is not restrictive: if instead $F_{\rm I} > R_{\rm I}$, the initial radiation-dominated phase would simply be absent. Thus, choosing $R_{\rm I} > F_{\rm I}$ captures the full range of possible cosmological histories for matter-like fluids with $w_\Phi < 1/3$.
	
	Our aim is to track the energy density of the radiation bath and use approximate solutions for $R(A)$ to identify the transitions between successive phases. At early times, the age of the universe is much shorter than the lifetime of $\Phi$, $\tau_\Phi = \Gamma_\Phi^{-1}$. Up to order-one factors, the age of the universe when the bath temperature is $T_{\rm I}$ is given by the inverse Hubble rate at that epoch. We are therefore in the regime $\Gamma_\Phi \ll H(T_{\rm I}) \simeq T_{\rm I}^2 / M_{\rm Pl}$. It is thus convenient to introduce the dimensionless parameter
	\begin{equation}
		\gamma_\Phi \equiv \frac{\sqrt{3}\,M_{\rm Pl}\,\Gamma_\Phi}{T_{\rm I}^2}\, .
	\end{equation}
	In what follows, we will make use of the hierarchy $\gamma_\Phi \ll 1$ to obtain approximate solutions.
	
	As long as the age of the universe remains shorter than $\tau_\Phi$, the comoving energy density of $\Phi$ stays approximately constant, so that the quantity defined in Eq.~\eqref{eq:Phi_comoving} satisfies $F(A) \simeq F_{\rm I}$. This approximation holds until the energy density of $\Phi$ eventually begins to decrease exponentially. We then track the corrections to the radiation comoving energy density in order to quantify its departure from the initial value $R_{\rm I}$ at $A_{\rm I}$. In particular, we identify the onset of the non-adiabatic phase as the moment when the contribution to radiation from $\Phi$ decays becomes dominant. All results derived below rely on the approximation $\gamma_\Phi \ll 1$, and we report only the leading terms in this expansion.
	
\begin{itemize}
	\item \textbf{RD$^\prime$ Phase.} In this phase, the radiation energy density in a comoving volume increases due to $\Phi$ decays, which add to the initial contribution $R_{\rm I}$, while $F(A) \approx F_{\rm I}$. We solve the differential equation in Eq.~\eqref{eq:bkg_comovingR}, using $w_R = 1/3$ as explained above and employing, for the Hubble parameter $H(A)$, the expression in Eq.~\eqref{eq:H_comoving} with only the radiation contribution retained. Implementing the appropriate initial conditions, we obtain
	\begin{equation}
    \begin{split}
		& \, R(A) \simeq R_{\rm I} + R^{\rm RD'}_{\rm decays}(A)	= \\ & = R_{\rm I} + \frac{\gamma_\Phi}{3 (1 - w_\Phi)} \frac{F_{\rm I}}{\sqrt{R_{\rm I}}}
		\left(A^{3(1 - w_\Phi)} - 1\right) \  .
    \end{split}
	\end{equation}
    This approximate solution is valid from the initial moment $A_{\rm I} = 1$ up to the onset of the adiabatic era dominated by $\Phi$, which begins at $A_{\rm A}$. The transition from RD$^\prime$ to $\Phi$D$_{\rm A}$ takes place when cosmological redshift brings the energy densities of $\Phi$ and radiation into equality. To determine the corresponding scale factor, we neglect the small contribution to $R(A)$ from $\Phi$ decays, obtaining
		\begin{equation}
			A_{\rm A} \simeq \left(\frac{R_{\rm I}}{F_{\rm I}}\right)^{1/(1 - 3 w_\Phi)} .
		\end{equation}
		At this moment, the thermal bath has temperature $T_{\rm A} \equiv T(A_{\rm A})$, which can be estimated from Eq.~\eqref{eq:R_comoving} by setting $A = A_{\rm A}$ and $R(A_{\rm A}) \simeq R_{\rm I}$.
		
		\item  ${\rm \boldsymbol\Phi \bf D_{\bf A}}$ \textbf{Phase}. During this phase the radiation bath becomes a sub-dominant component. We therefore solve the Boltzmann equation for the radiation energy density once more, now using the Hubble parameter appropriate for a period of $\Phi$ domination. Matching to the previous solution at the scale factor $A_{\rm A}$, we obtain the following expression for the comoving radiation energy density in the interval $A_{\rm A} < A < A_{\rm NA}$ 
		\begin{subequations}
			\begin{align}
				\label{eq:Ranalytical}
				&R(A) \simeq R_{\rm I} + R^{\rm RD'}_{\rm decays}(A_{\rm A}) + R^{\Phi{\rm D}_{\rm A}}_{\rm decays}(A)\,, \\
				\label{eq:RanalyticalNA}
				&R^{\Phi{\rm D}_{\rm A}}_{\rm decays}(A)
				 = \frac{2\gamma_\Phi}{5 - 3 w_\Phi}
				\sqrt{F_{\rm I}} \left[A^{(5 - 3 w_\Phi)/2}\right. \\
                &\nonumber\hspace{13em}\left.- A_{\rm A}^{(5 - 3 w_\Phi)/2}\right].
			\end{align}
			\label{eq:RanalyticalPhiA}
		\end{subequations}
		Three distinct components contribute to the comoving radiation energy density. The first, $R_{\rm I}$, is constant and represents the radiation already present at the initial time. The second, $R^{\rm RD'}_{\rm decays}(A_{\rm A})$, is also constant and accounts for the radiation produced by $\Phi$ decays during the RD$'$ phase; as noted above, the hierarchy $R^{\rm RD'}_{\rm decays}(A_{\rm A}) \ll R_{\rm I}$ always holds, so this term remains sub-dominant. The third contribution, $R^{\Phi{\rm D}_{\rm A}}_{\rm decays}(A)$, is the only one that depends explicitly on the scale factor and grows as time progresses. It arises from the decay products of $\Phi$ during the $\Phi$-dominated phase.
		
		As long as the radiation is mostly composed of the redshifted initial component, in other words as long as the right-hand side of Eq.~\eqref{eq:Ranalytical} is dominated by the term $R_{\rm I}$, the evolution remains adiabatic. In this regime, the bath temperature follows the usual radiation dominated scaling, $T \propto A^{-1}$. Once the contribution from the decay products of $\Phi$ becomes the largest term in the radiation energy density, the evolution enters the non-adiabatic phase. The semi analytical solution in Eq.~\eqref{eq:RanalyticalPhiA} allows us to identify the transition point between the adiabatic and the non-adiabatic stages of $\Phi$ domination. This transition occurs at the scale factor $A_{\rm NA}$ for which $R(A_{\rm NA}) = 2 R_{\rm I}$. Keeping only the leading term in the limit $\gamma_\Phi \ll 1$, we find
		\begin{align}\label{eq:A_NA}
			A_{\rm NA} \simeq \bigg[ \frac{5 - 3 w_\Phi}{2 \gamma_\Phi} \frac{R_{\rm I}}{\sqrt{F_{\rm I}}} \bigg]^{2/(5 - 3 w_\Phi)}\ .
		\end{align}
		The bath temperature at this moment, labelled as $T_{\rm NA} \equiv T(A_{\rm NA})$, is obtained from Eq.~\eqref{eq:R_comoving} evaluated at $A = A_{\rm NA}$ together with the condition $R(A_{\rm NA}) = 2 R_{\rm I}$.
		
		\item \textbf{${\rm \boldsymbol\Phi \bf D_{\bf NA}}$ Phase.} For values of the scale factor such that $A > A_{\rm NA}$, the universe enters the fully non-adiabatic regime, in which the radiation bath is dominated by the decay products of $\Phi$. This stage persists as long as $\Phi$ particles remain, that is until the age of the universe becomes comparable to the lifetime of $\Phi$. During this period the temperature no longer redshifts adiabatically as $T \propto A^{-1}$. Instead, its dependence on the scale factor follows from Eq.~\eqref{eq:R_comoving}, with $R(A)$ well approximated by the leading term in Eq.~\eqref{eq:RanalyticalNA}.
		
		Once all $\Phi$ particles have decayed, the universe enters the final radiation dominated era. The value of the scale factor $A_{\rm R, dec}$ at this moment is identified by requiring that the age of the universe equals the lifetime of $\Phi$. In a $\Phi$ dominated universe the scale factor evolves as $a(t) \propto t^{2/[3(1+w_\Phi)]}$, which implies $H(t) = 2/[3(1+w_\Phi)\, t]$. We therefore determine the reheating moment from the condition $H(A_{\rm R, dec}) = 2\,\Gamma_\Phi/[3(1+w_\Phi)]$. This gives
		\begin{align}
			A_{\rm R, dec} \simeq 
			\left[ \frac{3(1 + w_\Phi)}{2 \gamma_\Phi}\, \sqrt{F_{\rm I}} \right]^{2/[3(1 + w_\Phi)]}\, .
			\label{eq:AR}
		\end{align}
		An alternative way to identify the reheating moment is to solve the differential equation for $\Phi$ in Eq.~\eqref{eq:bkg_comovingF}. With our choice of initial conditions, the corresponding integral receives contributions from both the radiation dominated phase and the subsequent $\Phi$ dominated phase. In each regime we substitute the appropriate expression for the Hubble parameter and approximate the comoving densities as constant, consistent with the limit $\gamma_\Phi \ll 1$. Evaluating the integral under these assumptions yields
		\begin{equation}
        \begin{split}
			&\qquad F(A) = \, F_{\rm I} \exp\!\left[ -\,\Gamma_\Phi \int_1^{A} \frac{d\log A'}{H(A')} \right]
			 \\
             &\qquad\qquad\,\,\,\simeq
			F_{\rm I} \exp\!\left[ - \frac{\gamma_\Phi}{2}\frac{A_{\rm A}^2 - 1}{\sqrt{R_{\rm I}}}
			-\right. \\
            &\hspace{8em} \left.\frac{2\gamma_\Phi}{3(1+w_\Phi)} \frac{ A^{\frac{3(1+w_\Phi)}{2}} - A_{\rm A}^{\frac{3(1+w_\Phi)}{2}} }{\sqrt{F_{\rm I}}} \right].
        \end{split} 
		\end{equation}
		Defining the reheating scale factor $A_{\rm R}$ as the moment when $F(A_{\rm R, dec}) = F_{\rm I}/e$, and neglecting subleading terms in the exponent, we recover the same expression as in Eq.~\eqref{eq:AR}. The corresponding reheating temperature follows from the same condition used to identify $A_{\rm R}$, combined with the Friedmann equation for a radiation dominated universe, and explicitly reads
		\begin{equation}\label{eq:TR}
		    \qquad\,\,\frac{2\,\Gamma_\Phi}{3(1+w_\Phi)} = \frac{1}{\sqrt{3} M_{\rm Pl}} \left( \frac{\pi^2}{30} g_{\star}(T_{\rm R, dec})\, T_{\rm R, dec}^4 \right)^{1/2} \ .
		\end{equation}
	
        \end{itemize}
	
	To summarize, a cosmological history with $w_\Phi < 1/3$ is fully specified by four parameters, $(w_\Phi, R_{\rm I}, F_{\rm I}, \Gamma_\Phi)$. Once $R_{\rm I}$, $F_{\rm I}$, and $w_\Phi$ are fixed, one may trade $\Gamma_\Phi$ for the reheating temperature $T_{\rm R, dec}$. It is then necessary to verify that this choice is compatible with the onset of the standard radiation dominated era, in particular that $T_{\rm R, dec} > 5~\mathrm{MeV}$.
	
\subsection{Semi-analytical solutions for kination-like $\Phi$}
\label{supp:cosmo_w<1/3}
	
We now consider the case of a kination-like fluid with $w_{\Phi} > 1/3$. In this regime the exotic component $\Phi$ redshifts faster than radiation. As shown in the right panel of Fig.~\ref{fig:histories}, such a component can affect the cosmological evolution only if $F_{\rm I} > R_{\rm I}$; otherwise the universe remains radiation dominated at all times and the standard cosmological history is recovered. When the initial conditions satisfy $F_{\rm I} > R_{\rm I}$, the universe undergoes a phase of $\Phi$ domination before eventually returning to radiation domination. As discussed in the main text and illustrated in Fig.~\ref{fig:histories}, there are two distinct scenarios through which radiation can later become dominant. We discuss these two cases separately.
	
\begin{enumerate}
		
	\item[i)] The first case corresponds to a $\Phi$ particle that is sufficiently long-lived that it redshifts away before its decay products have any tangible impact. This includes the situation in which $\Phi$ is absolutely stable. This case is illustrated by the dashed lines in Fig.~\ref{fig:histories}. In this regime, radiation eventually becomes dominant purely because of the different redshifting behaviors, and this occurs at the scale factor
		\begin{align}
			A_{\rm R,rsh} \simeq \left(\frac{F_{\rm I}}{R_{\rm I}}\right)^{1/(3w_\Phi - 1)}\ .
			\label{eq:ARrsh}
		\end{align}
	The corresponding temperature of the radiation bath, $T_{\rm R,rsh}$, can be estimated from Eq.~\eqref{eq:R_comoving} by evaluating it at $A = A_{\rm R,rsh}$ and $R(A_{\rm A}) \simeq R_{\rm I}$. Therefore, stable or sufficiently long-lived $\Phi$ particles lead to a simple cosmological history consisting of two eras. First, there is an adiabatic phase dominated by the $\Phi$ field, which we denote as $\rm \Phi D$ (without the subscript A to emphasize the absence of any non-adiabatic stage). This is followed by the standard RD era. These two phases are separated by the scale factor $A_{\rm R,rsh}$ given in Eq.~\eqref{eq:ARrsh}.
		
	\item[ii)] The second case corresponds to a $\Phi$ particle that is sufficiently short-lived to decay while it still dominates the energy budget of the universe. This situation is illustrated by the solid lines in Fig.~\ref{fig:histories}. Up to order one factors, this scenario takes place when $\Gamma_\Phi \gtrsim H(A_{\rm R,rsh})$. A decaying $\Phi$ leads to a cosmological history analogous to the case with $w_\Phi < 1/3$, but without the initial radiation dominated phase RD$'$. The evolution therefore begins with a $\rm \Phi D_{A}$ phase, which ends at $A_{\rm NA}$ given in Eq.~\eqref{eq:A_NA}, when a non-adiabatic $\rm \Phi D_{NA}$ era starts. This era persists until the $\Phi$ particles decay, and the corresponding scale factor $A_{\rm R,dec}$ can be identified by using the expression in Eq.~\eqref{eq:AR}. The reheat temperature of the radiation bath, $T_{\rm R,dec}$, is given by an expression identical to Eq.~\eqref{eq:TR}. After this stage, the universe transitions to the standard radiation dominated era.
		
\end{enumerate}
	
In summary, for $w_\Phi > 1/3$, a given cosmological history is fully specified by four parameters, $(w_\Phi, R_{\rm I}, F_{\rm I}, \Gamma_\Phi)$, when $\Gamma_\Phi \gtrsim H(A_{\rm R,rsh})$. Otherwise only three parameters, $(w_\Phi, R_{\rm I}, F_{\rm I})$, are required. Once $R_{\rm I}$, $F_{\rm I}$, and $w_\Phi$ are fixed, one may, when relevant, trade $\Gamma_\Phi$ for the reheating temperature $T_{\rm R,dec}$. In addition, one must ensure that $T_{\rm R,dec} > 5~\mathrm{MeV}$ or $T_{\rm R,rsh} > 5~\mathrm{MeV}$, depending on the scenario.

\subsection{Entropy injection and dilution factor}
	
An important quantity during non-adiabatic cosmological histories is the amount of entropy injected into a comoving volume between the onset of the non-adiabatic phase and a later moment. For a scale factor $A$ such that the universe is still in the non-adiabatic era, that is for $A_{\rm NA} < A < A_{\rm R, dec}$, the ratio between these entropies is defined as
\begin{align}
\xi_S(A) \equiv \frac{S(A)}{S(A_{\rm NA})} = \frac{g_{\star s}(T(A))\, T(A)^{3}\, A^{3}} {g_{\star s}(T_{\rm NA})\, T_{\rm NA}^{3}\, A_{\rm NA}^{3}} \ ,
\label{eq:Ddef}
\end{align}
where $g_{\star s}(T)$ denotes the effective number of entropic degrees of freedom and $T(A)$ is the bath temperature at that moment.
To understand how $\xi_S(A)$ scales with the bath temperature, we can use the approximate solutions derived in the previous subsections, which yields
\begin{align}
		\xi_S(T) \simeq 
		\frac{g_{\star s}(T(A))}{g_{\star s}(T_{\rm NA})}
		\left[\frac{30}{\pi^2} 
		\frac{2 R_{\rm I} T_{\rm I}^4}{g_\star(T_{\rm NA})} \right]^{-3/4}
		T(A)^{3} \, A^{3} \, .
\end{align}
This implies the scaling $\xi_S(T) \propto T^{\,3 - \frac{8}{1+w_\Phi}}$. For the specific case of a matter-dominated era, $w_\Phi = 0$, we recover the well-known scaling $\xi_S(T) \propto T^{-5}$.
	
A particularly important role in characterizing a cosmological history is played by the total amount of entropy injected during the non-adiabatic phase. We define this quantity as $D_{\rm R} \equiv \xi_S(A_{\rm R, dec})$ and refer to it as the \textit{dilution factor}, since it quantifies the dilution experienced by comoving number densities as a consequence of the entropy injection. We evaluate $D_{\rm R}$ using the general definition in Eq.~\eqref{eq:Ddef}, providing semi-analytical estimates in the limit $g_{\star s}(T) \simeq g_\star(T) \simeq \text{const}$. Within these approximations, we obtain
	\begin{align}
		D_{\rm R} \simeq 
		2^{\frac{3 w_\Phi - 5}{4(1+w_\Phi)}}\;
		\big[3(1+w_\Phi)\big]^{\frac{1-3 w_\Phi}{2 (1 + w_\Phi)}} \,
		R_{\rm I}^{-3/4} \,
		F_{\rm I}^{\frac{1}{1+w_\Phi}} \,
		\gamma_\Phi^{\frac{3 w_\Phi - 1}{2(1+w_\Phi)}}.
		\label{eq:fullD}
	\end{align}
	Two well-known limits are readily recovered:
	$D_{\rm R} \simeq R_{\rm I}^{-3/4} F_{\rm I}\,\gamma_\Phi^{-1/2}$ for $w_\Phi = 0$, and 
	$D_{\rm R} \simeq R_{\rm I}^{-3/4} F_{\rm I}^{1/2}\,\gamma_\Phi^{1/2}$ for $w_\Phi = 1$. 
	Note that the dependence on $R_{\rm I}$ appears only through an overall multiplicative factor, independently of $w_\Phi$.
	
	The expression in Eq.~\eqref{eq:fullD} helps identify the regions of parameter space that lead to a similar non-adiabatic era. More specifically, for each value of $w_\Phi$, the isocontours of constant $D_{\rm R}$ display different shapes. If $D_{\rm R} \sim \gamma_\Phi^{a} F_{\rm I}^{b}$, these isocontours satisfy $F_{\rm I} \sim \gamma_\Phi^{-a/b}$, while a trajectory perpendicular to them corresponds to $F_{\rm I} \sim \gamma_\Phi^{b/a}$.
	
	We turn to DM production via freeze-in in the next section. As an immediate application of the dilution factor, we anticipate that any comoving number density of DM particles produced before the onset of a non-adiabatic era will be diluted by a multiplicative factor inversely proportional to $D_{\rm R}$. If DM production takes place during a non-adiabatic era, the dilution is only partial, since only the entropy injected after the production moment affects the comoving number density. This is the reason why we introduced in the main text the mother-particle–mass–dependent dilution factor $D(M)$. In general, we define the partial dilution factor $D(M) \equiv S(A_{\rm R, dec})/S(A_M)$, with $A_M = a_M/a_{\rm I}$ the value of the dimensionless scale factor for which $T(A_M) = M$, as the amount of entropy dumped in a comoving volume between the moment when the bath temperature was $M$ and the end of the non-adiabatic era.
	
	
\section{Freeze-in production in general cosmologies}
\label{supp:PSD}
	
In this Appendix, we provide semi-analytical results for DM produced via freeze-in during one of the cosmological eras identified in Fig.~\ref{fig:histories}. We begin by obtaining a formal solution to the Boltzmann equation and show how to derive the analytical shape of the phase-space distribution (PSD). Once the PSD is known, we explain how to extract its second moment and, from it, the lower bound on the DM mass by mapping the result onto the warm dark matter (WDM) bounds. We then specialize these results to the various cosmological phases in which DM production may occur.
	
	The Boltzmann equation in Eq.~\eqref{eq:Boltzmann}, after trading the time variable for the reduced scale factor $A = a/a_{\rm I}$, becomes
	\begin{align}
		\frac{df_\chi(A,p)}{d\log A} = \frac{{\cal C}(T(A),p(A))}{H(A)} \, .
	\end{align}
	The formal solution from $A_{\rm I}=1$ to a final value $A_{\rm F}$ such that $T(A_{\rm F}) \ll 1~\text{MeV}$ is
	\begin{align}
		f_\chi(A_{\rm F},p) 
		= \int_{0}^{\log A_{\rm F}} d\log A \, 
		\frac{{\cal C}(T(A),p(A))}{H(A)} \, .
		\label{eq:BoltzSol}
	\end{align}
	Assuming that DM production takes place during a specific cosmological era, the integrand is dominated by a well defined epoch in the expansion history. We therefore perform the integration using the scalings of two fundamental quantities, the Hubble parameter and the bath temperature, parameterized as $H \propto a^{m}$ and $T \propto a^{\ell}$. For example, up to $g_\star(T)$ corrections, during standard radiation domination the two exponents are $(m,\ell)=(-2,-1)$.
	
\subsection{Production via two-body decays}
	
We specialize to the production channel considered in this work: freeze-in through two-body decays of a thermal bath particle ${\cal B}_1$, focusing on single production, ${\cal B}_1 \to {\cal B}_2 \chi$. For freeze-in, the case of double production is equivalent to single production up to an overall factor of two in the final relic density. We further assume Maxwell–Boltzmann statistics for all particles and a mass hierarchy $m_{{\cal B}_2}, m_\chi \ll m_{{\cal B}_1} = M$. Under these assumptions the collision term is analytical and, neglecting overall multiplicative constants, can be written as~\cite{DEramo:2020gpr}
	\begin{align}\label{eq:Collision_gen}
		{\cal C}_{{\cal B}_1 \to {\cal B}_2 \chi}(T, p) \propto \frac{T\, e^{-(E + E_2^-)/T}}{p\, E}\, |{\cal M}_{{\cal B}_1 \to {\cal B}_2 \chi}|^2 \ .
	\end{align}
	Here, $E$ and $p$ denote the energy and the magnitude of the spatial momentum of the DM particle, respectively. The squared matrix element for this process, $|{\cal M}_{{\cal B}_1 \to {\cal B}_2 \chi}|^2$, is independent of the kinematics because two-body decays produce monochromatic final states. Moreover, it is proportional to the decay width of ${\cal B}_1$ and, within our assumptions about the mass spectrum, it explicitly reads $|{\cal M}_{{\cal B}_1 \to {\cal B}_2 \chi}|^2 = 16 \pi\, \Gamma_{{\cal B}_1 \to {\cal B}_2 \chi}\, M$. Finally, $E_2^-$ is the minimum energy of the bath particle in the final state ${\cal B}_2$, and it evaluates to $E_2^- / T \simeq M^2 / (4 p T)$. We assume DM to be relativistic at production, $E \simeq p$. Putting everything together, and including the appropriate multiplicative factor, one recovers the expression in Eq.~\eqref{eq:CollisionTerm2Bdec} of the main text. For our analysis of DM mass bounds, the overall factor does not affect the shape of the DM distribution and becomes relevant only when we impose the relic density condition.
	
	It is convenient to employ the comoving momentum defined as
	\begin{align}\label{eq:comoving_q}
		q \equiv \frac{p}{M}\frac{a}{a_M} \equiv \frac{p}{M} \times \widetilde{A} \ ,
	\end{align}
	where $a_M$ is the value of the scale factor for which the thermal bath temperature satisfies $T(a_M)=M$, and $\widetilde A \equiv a/a_M = A/A_M$ with $A_M = a_M/a_{\rm I}$. The Hubble rate and the bath temperature can then be written as
	\begin{subequations}\label{eq:HT(A)}
		\begin{align}
			H &= H_M \, \widetilde A^{m}\ ,\\
			T &= M\, \widetilde A^{\ell}\ .
		\end{align}
	\end{subequations}
	where $H_M$ is the value of the Hubble parameter when the thermal bath has a temperature equal to $M$. Replacing these definitions back into Eq.~\eqref{eq:Collision_gen}, we obtain the scaling of the collision term
	\begin{align}
		{\cal C}_{{\cal B}_1 \to {\cal B}_2 \chi}(\widetilde A, q) \propto 
		\frac{\widetilde A^{\ell+2}}{q^2}\,
		\exp\!\bigg[ - q\, \widetilde A^{-\ell-1} - \frac{1}{4q}\, \widetilde A^{1-\ell} \bigg]\ .
	\end{align}
	We then insert this expression into the formal solution given in Eq.~\eqref{eq:BoltzSol}, noting that we have changed integration variable from $A$ to $\widetilde A$, which is more suitable for the analysis of DM production via freeze-in. This yields the following shape for the DM distribution,
	\begin{align}\label{eq:f_analytical}
		f_\chi(\widetilde A_{\rm F}, q) \propto 
		\frac{1}{q^2}
		\int_{\widetilde A_{\rm I}}^{\widetilde A_{\rm F}}
		d\widetilde A\, 
		\widetilde A^{\ell - m + 1}\,
		\exp\!\bigg[ -\frac{q}{\widetilde A^{\ell+1}} - \frac{\widetilde A^{1-\ell}}{4q}\,\bigg].
	\end{align}
	This expression coincides with the one given in Eq.~\eqref{eq:PSD_an} of the main text. It is useful because it allows us to evaluate the second moment of the PSD in comoving variables, $\sigma_q$, by first performing the integration over $q$ before integrating over the reduced scale factor $\widetilde A$. We find the explicit result in terms of modified Bessel functions,
	\begin{align}\label{eq:sigmaq}
		\sigma_q = \sqrt{\frac{\int_0^\infty dq\, q^4 f_\chi(q)}{\int_0^\infty dq\, q^2 f_\chi(q)}} = \frac{1}{2}\sqrt{\frac{\int_{\widetilde A_{\rm I}}^{\widetilde A_{\rm F}} d\widetilde A\, \widetilde A^{\ell - m + 4} K_3(\widetilde A^{-\ell})}{\int_{\widetilde A_{\rm I}}^{\widetilde A_{\rm F}} d\widetilde A\, \widetilde A^{\ell - m + 2} K_1(\widetilde A^{-\ell})}} \ ,
	\end{align}
	which reproduces the result given in Eq.~\eqref{eq:sigmaq_an} of the main text.
	
\subsection{Mapping WDM mass bounds onto freeze-in production: general strategy}
	
	Once the PSD second moment is obtained, it is straightforward to map it into a lower bound on the DM mass. This follows from requiring that the r.m.s. DM velocity be smaller than that of WDM when the latter is set to its minimum allowed mass. We start from the general expression (see, e.g., App.~A of Ref.~\cite{DEramo:2025jsb})
	\begin{align}
		\, W_{\rm WDM}(t_0)& \equiv \frac{\sqrt{\langle p_0^2\rangle_{\rm WDM}}}{m_{\rm WDM}} = \frac{\sigma_q^{\rm WDM} T_{\rm WDM}}{m_{\rm WDM}} \\ &=\nonumber 1.2 \times 10^{-8} \left( \frac{\sigma_q^{\rm WDM}}{3.6} \right) \left( \frac{6\,{\rm keV}}{m_{\rm WDM}} \right)^{4/3}\\
        &\nonumber\hspace{4em}\times\left( \frac{2}{g_{\rm WDM}} \frac{3/4}{g_{\rm eff}} \frac{\Omega_{\rm WDM} h^2}{0.12} \right)^{1/3}\, .
	\end{align}
	Here $\sigma_q^{\rm WDM}$ is the dimensionless second moment of the WDM distribution. In the last equality we normalized the expression by assuming that the correct relic abundance is reproduced and specializing to the fermionic WDM case. We further normalized it to $m_{\rm WDM}=6\,{\rm keV}$, which we take as the benchmark value for the WDM mass bound.
	
	The lower bound on the DM mass produced via freeze-in is obtained by imposing $W_\chi(t_0) \equiv \sqrt{\langle p_0^2\rangle}/m_\chi < W^{\rm max}_{\rm WDM}(t_0)$. The maximum allowed WDM r.m.s.\ velocity corresponds to setting the WDM mass to its minimum value allowed by structure-formation bounds, which leads to
	\begin{equation}\label{eq:m_min_gen}
    \begin{split}
    m_\chi^{\rm min} &=  \frac{\sqrt{\langle p_0^2\rangle}}{W^{\rm max}_{\rm WDM}(t_0)}\\
		&= \frac{\sigma_q D(M)^{-1/3} T_\chi(t_0)}{W^{\rm max}_{\rm WDM}(t_0)}
		= \frac{\Sigma\, T_\chi(t_0)}{W^{\rm max}_{\rm WDM}(t_0)}\\ &=  19\, {\rm keV}
		\left(\frac{m_{\rm WDM}}{6\ {\rm keV}}\right)^{4/3}
		\left(\frac{\Sigma}{3}\right)
		\left(\frac{104.4}{g_\star(M)}\right)^{1/3}\,,
    \end{split}
	\end{equation}
	where we have introduced the effective DM temperature $T_\chi(t_0)=T_0\,(g_{\star s}(T_0)/g_{\star s}(M))^{1/3}$ associated with the production temperature $M$ and the current CMB temperature $T_0 \simeq 2.35 \times 10^{-4}\,{\rm eV}$. The dilution-corrected warmness parameter $\Sigma \equiv \sigma_q D(M)^{-1/3}$ includes two effects. First, it scales with the second moment of the PSD, $\sigma_q$, which depends on the production channel (two-body decays in our case) and on the cosmological era at production. Second, it incorporates the impact of entropy injection from $\Phi$ decays during the non-adiabatic era, encoded in the dilution factor $D(M)$, which quantifies the additional redshift of DM momenta from $T=M$ to today and is highly sensitive to the cosmological history both \textit{during} and \textit{after} production.
	
	Deriving analytical results directly from the PSD in Eq.~\eqref{eq:f_analytical} is not always feasible. The equation of state of the exotic fluid $\Phi$, which fixes the exponents $(\ell,m)$, can introduce nontrivial dependencies in the integrand. Additional subtleties arise from the specific cosmological phase during which freeze-in occurs, and fully analytic solutions become particularly challenging in a non-adiabatic era. For this reason, in the next two subsections we focus on DM production during adiabatic epochs and distinguish between the cases without and with post-production dilution.
	
\subsection{DM production during adiabatic phases without subsequent dilution}
	
For freeze-in production during adiabatic phases, analytical results for the present-day PSD and $\sigma_q$ can be obtained by taking the limits $\widetilde A_{\rm I}\!\to 0$ and $\widetilde A_{\rm F}\!\to \infty$ in Eqs.~\eqref{eq:f_analytical} and \eqref{eq:sigmaq}. This is justified because, as long as entropy is conserved during the production window, there is no dilution: the DM abundance simply scales with the expansion of the universe while the momentum distribution redshifts in the usual way. Consequently, the duration of the cosmological phase has no impact on the final abundance or on the momentum distribution. The analytic results obtained in these frameworks are summarized in Tab.~\ref{tab:analytics}.
	
\begin{table}[t]
		\centering
		\begin{tabular}{c||c|c||c|c}
			Cosmology & $T \propto a^{\ell}$ & $H \propto a^{m}$ & $f_\chi(q)$ & $\sigma_q$ \\\hline
			RD/RD${}'$        & $a^{-1}$ & $a^{-2}$ & $ q^{-1/2} e^{-q}$ &
			$\sqrt{35}/2$ \\
			$\Phi{\rm D_A}$/\,$\Phi{\rm D}$ & $a^{-4}$ &
			$a^{-3(1+w_{\Phi})/2}$ & 
			$ q^{\frac{3(w_\Phi-1)}{4}} e^{-q}$ &
			$\sqrt{\frac{\Gamma \left(\frac{3 w_\Phi+17}{4}\right)}
				{\Gamma \left(\frac{3w_\Phi+9}{4}\right)}}$
\end{tabular}
\caption{Shape of the DM PSD $f_\chi(q)=f_\chi(\widetilde A_{\rm F}\!\to\!\infty,q)$ and its second moment $\sigma_q$ for freeze-in via two-body decays in different cosmological backgrounds, assuming production occurs entirely within a single era. The first column identifies the epoch during which DM is produced. Production during a non-adiabatic era is not included because the limit $\widetilde A_{\rm F}\!\to\!\infty$ cannot be taken in that case, as discussed in the main text.}
\label{tab:analytics}
\end{table}
	
The first two cases we discuss correspond to DM production, in the language of Fig.~\ref{fig:histories}, during the RD or $\Phi$D eras. Since no subsequent non-adiabatic phase is present, the dilution factor is $D(M)=1$, and the second moment of the PSD computed in terms of comoving or physical momentum satisfies $\Sigma=\sigma_q(w_\Phi)$. The corresponding values for each phase are listed in Tab.~\ref{tab:analytics}, and we report them here for convenience
\begin{subequations}\label{eq:sigma_RD_PHID}
	\begin{align}\label{eq:sigmaRD}
		{\rm RD:}\qquad & \Sigma(w_\Phi) = \frac{\sqrt{35}}{2}\,, \\
		{\rm \Phi D:}\qquad & 
		\Sigma(w_\Phi) = \sqrt{\frac{\Gamma \left(\frac{3 w_\Phi+17}{4}\right)}
			{\Gamma \left(\frac{3w_\Phi+9}{4}\right)}}\,.
		\label{eq:sigmaPhiD}
	\end{align}
\end{subequations}
The lower bound on the DM mass then follows from Eq.~\eqref{eq:m_min_gen}, yielding
\begin{subequations}\label{eq:bound_RD_PHID}
	\begin{align}
    {\rm RD:}\quad 
			m_\chi^{\rm min} &\approx
			2.96\,\frac{T_\chi(t_0)}{W_{\rm WDM}(t_0)}\\
			&\nonumber=19\,{\rm keV}\left(\frac{m_{\rm WDM}}{6~{\rm keV}}\right)^{4/3}\left(\frac{104.4}{g_\star(M)}\right)^{1/3}, \\
    {\rm \Phi D:}\quad 
			m_\chi^{\rm min} &\approx (1.94 - 3.46)\,
			\frac{T_\chi(t_0)}{W_{\rm WDM}(t_0)}\\
			&\nonumber=(12 - 22)\,{\rm keV}\left(\frac{m_{\rm WDM}}{6~{\rm keV}}\right)^{4/3}\left(\frac{104.4}{g_\star(M)}\right)^{1/3}.
	\end{align}
\end{subequations}
For production during the $\Phi$D era, the quoted range reflects varying the parameter $w_\Phi$ within $(-1,1)$.
	
\subsection{DM production during adiabatic phases with subsequent dilution}
\label{supp:PSD_analytics}
	
We now consider the situation in which a non-adiabatic phase follows the DM production epoch. Referring to Fig.~\ref{fig:histories}, this corresponds to freeze-in during the RD$'$ or $\Phi{\rm D_A}$ eras. In these cases, we must account for the effect of the dilution factor $D(M)$. By construction, this factor is bounded from below, $D(M)\geq 1$, but it is also bounded from above: an excessive dilution would require a very large coupling to reproduce the relic density, eventually invalidating the freeze-in framework. In practice, we must therefore ensure that DM never thermalizes with the radiation bath.
	
To verify this condition, we introduce the DM yield $Y_\chi \equiv n_\chi/s$. The relic abundance requirement implies $Y_\chi(A_0) = 0.44\,{\rm eV}/m_\chi$ today. To ensure that DM never thermalizes, we require the yield to remain well below its equilibrium value at all times. We impose condition $Y_\chi(A) < 0.1\,Y_\chi^{\rm eq}(A)$, which translates into the constraint $Y_\chi(A_0)\,D(M) < Y_\chi^{\rm eq}$.
	
The minimum conceivable value of the equilibrium yield is obtained for a massless species with one internal degree of freedom at temperatures well above the weak scale, giving $Y_\chi^{\rm eq} \simeq 0.0022$. Substituting this into the freeze-in condition yields a constraint on the dilution factor
\begin{align}
	1 \leq D(M) \lesssim \max\!\left[1,\, \frac{m_\chi}{2~{\rm keV}}\right]\,.
\end{align}
	
If $D(M)=1$, the minimum allowed mass $m_\chi^{\rm min}$ coincides with Eqs.~\eqref{eq:bound_RD_PHID}. Otherwise, we return to Eq.~\eqref{eq:m_min_gen}, incorporate the maximum dilution factor, and derive an equation that must be solved consistently for $m_\chi^{\rm min}$
\begin{multline}
        m_\chi^{\rm min}= 19\ {\rm keV}
	\left(\frac{m_\chi^{\rm min}}{2\ {\rm keV}}\right)^{-1/3}
	\left(\frac{m_{\rm WDM}}{6\ {\rm keV}}\right)^{4/3}\\\times 
	\left(\frac{\sigma_q}{3}\right)
	\left(\frac{104.4}{g_\star(M)}\right)^{1/3}\,.
\end{multline}
The resulting bound is
\begin{align}
	m_\chi^{\rm min}= 11\ {\rm keV}
	\left(\frac{m_{\rm WDM}}{6\ {\rm keV}}\right)
	\left(\frac{\sigma_q}{3}\right)^{3/4}
	\left(\frac{104.4}{g_\star(M)}\right)^{1/4}\,.
\end{align}
Using the analytical expressions for $\sigma_q$ from Tab.~\ref{tab:analytics}, we obtain the following ranges for the DM mass bounds
\begin{subequations}\label{eq:bound_RDp_PHIDA}
	\begin{align}
		{\rm RD':}\qquad &
		m_\chi^{\rm min} \approx 11\ {\rm keV}
		\left(\frac{m_{\rm WDM}}{6\ {\rm keV}}\right)
		\left(\frac{104.4}{g_\star(M)}\right)^{1/4}, \\
		{\rm \Phi D_A:}\qquad &
		m_\chi^{\rm min} \approx (7.8 - 12)\ {\rm keV}
		\left(\frac{m_{\rm WDM}}{6\ {\rm keV}}\right)
		\left(\frac{104.4}{g_\star(M)}\right)^{1/4}
	\end{align}
\end{subequations}
Moreover, the corresponding values of $\Sigma$ for these two phases are
\begin{subequations}\label{eq:sigma_RDp_PHIDA}
	\begin{align}\label{eq:sigmaRDp}
		{\rm RD':}\qquad &
		\Sigma(w_\Phi,m_\chi) = 
		\frac{\sqrt{35}}{2}\,
		\max\!\left[1,\, \frac{m_\chi}{2\ {\rm keV}}\right], \\
		{\rm \Phi D_A:}\qquad &
		\Sigma(w_\Phi,m_\chi) =
		\sqrt{\frac{\Gamma\!\left(\frac{3 w_\Phi+17}{4}\right)}
			{\Gamma\!\left(\frac{3 w_\Phi+9}{4}\right)}}\,
		\max\!\left[1,\, \frac{m_\chi}{2\ {\rm keV}}\right].
	\end{align}\label{eq:sigmaPHIDA}
\end{subequations}
    
\bibliographystyle{apsrev4-1}
\bibliography{ref} 

\end{document}